\documentstyle[editedvolume]{crckapb}

\input{psfig}
\begin{opening}
\title{A Multi-frequency campaign on the $\gamma$-ray loud blazar W~Comae}
\author{M. Maisack$^1$}
\author{K. Mannheim$^2$}
\author{R.D. Geckeler$^1$}
\author{M. Hillas$^3$}
\author{S. Katajainen$^4$}
\author{F. Marshall$^5$}
\author{D. Petry$^6$}
\author{C. v. Montigny$^7$}
\author{G. Paubert$^8$}
\author{J. Rose$^3$}
\author{A. Sillan\-p\"a\"a$^4$}
\author{L.O. Takalo$^4$}
\author{H. Ter\"asranta$^9$}
\institute{$^1$ Universit\"at T\"ubingen, Germany\\
$^2$ Universit\"at G\"ottingen, Germany\\
$^3$ Leeds University, UK\\
$^4$ Tuorla Observatory, Piikki\"o, Finland\\
$^5$ Goddard Space Flight Center, Greenbelt, USA\\
$^6$ Max-Planck-Institut f\"ur Physik, M\"unchen, Germany\\
$^7$ Landessternwarte Heidelberg, Germany\\
$^8$ IRAM, Granada, Spain\\
$^9$ Mets\"ahovi Radio Research Station, Kylm\"al\"a, Finland\\
}
\end{opening}

\runningtitle{A Multi-frequency campaign on the blazar W~Comae}

\begin{document}

\begin{abstract}

We report preliminary results of a multi-frequency campaign on the 
TeV candidate blazar W~Comae ($z$=0.102). Flux limits by 
Whipple and HEGRA show that the TeV flux must be considerably
below an $E^{-2}$ extrapolation of the EGRET flux.
In the framework of proton-initiated cascade models,
this seems to imply weak $\gamma$-ray attenuation due to pair production
in collisions with low-energy photons of the extragalactic
infrared background.
In a simple SSC model, the $\gamma$-ray spectrum cuts off intrinsically
just below TeV.

\end{abstract}

\section{Introduction}

The window of $\gamma$-ray astronomy has been pushed wide open 
by the CGRO satellite and ground-based Cherenkov light 
detectors such as Whipple and HEGRA. In the extragalactic domain, 
EGRET has detected (as of October 1996) 68 blazar-type 
AGN (Kanbach 1996) at energies of 100~MeV to several GeV.
Whipple and HEGRA have so far detected two (three)
blazars at TeV energies.
Detections at TeV energies are expected to be rare, since the 
extragalactic IR background (EIB) 
attenuates $\gamma$-rays by photon-photon pair 
production, making only the detection of nearby 
AGN feasible. Gamma-rays
of energy $E$  from sources with
redshift $z\ll 1$
preferentially produce pairs in collisions with isotropic
near-IR photons of energy
$\epsilon\approx {0.5 {\rm eV}/ E[{\rm TeV}]}$.
Due to a greater path length,
the cutoff in the $\gamma$-ray spectrum occurs at lower energies 
for higher redshift sources.  
Exact predictions of the expected cutoff 
are not possible yet, since the shape of 
the EIB -- in contrast 
to the 3K microwave background -- is not well known. Recent predictions of 
the EIB based on cold and hot dark matter galaxy formation models 
(MacMinn and Primack 1996) suggest that nearby blazars 
with $z$ 
up to $\sim$ 0.1 can be observed at TeV energies and
therefore represent suitable probes  
of the EIB (Mannheim et al. 1996). 

TeV observations are also rewarding for tests of blazar emission models. 
While most of the currently popular models invoke emission from a plasma 
flowing relativistically from the center of the AGN at small angles towards 
the observer, the origin of the soft photons scattered into the
$\gamma$-ray regime by relativistic particles, the distance from the AGN 
center, and the nature of the relativistic particles in the jet
are still a matter of debate. A model that naturally predicts TeV 
emission was proposed by Mannheim (1993), the so-called proton initiated 
cascade model (PIC). Here, 
the jet thrust resides in protons (contrary to other models which use 
electrons or positrons), and
protons and electrons
are accelerated in situ, thus avoiding the problem of photon 
drag in the vicinity of the central engine and the associated over-production
of soft X-rays (Sikora and Madejski 1996).

\begin{table}[t]\centering
\begin{tabular}[c]{llll}
\hline 
Frequency & Instrument & Date & Flux		\\
\hline
Radio  & Mets\"ahovi & continuous & $S_{22}=0.67\pm0.03$~Jy (27.2)\\
& (22,37 GHz) & & $S_{37}=0.50\pm0.08$~Jy (3.3)\\
\hline
mm    & Pic du Midi & Mar.10 & $S_{243}=0.41\pm0.04$~Jy\\
\hline
UBVRI & 0.8m McDonald & Feb.23-Mar.7 & $m_{\rm V}=14.40\pm0.08$ (26.2,29.2,2.3)\\
V & 1.0m Tuorla & Jan 28-Mar.3 & $m_{\rm V}=13.90\pm0.05$ (6.2,10.2)\\
\hline
UV & IUE &  Feb.08 	&  
$F_{1400}=(3.2\pm0.3)\times 10^{-14}$~${\rm ergs
\over cm^2~s~\AA}$	\\
\hline
X-rays & XTE & Feb.17. & $F_{2-10}<(3.5\pm 1.5)\times 10^{-5}$~Crab \\
\hline
100~MeV & EGRET & Feb.20-Mar.5 & 
$F_{>100}=(19.2\pm 6.0)\times 10^{-8}~{\rm photons\over
cm^2~s}$\\
\hline
TeV &	Whipple & Jan-Feb  & $F_{>0.35}^{3\sigma} < 7.7\times10^{-12}~{\rm photons\over
cm^2~s}$\\
 & HEGRA & Feb. &  $F_{>1.0}^{3\sigma} < 1.05\times10^{-11}~{\rm photons\over cm^2~s}$\\
&&13,16-19,26,27 &$F_{>1.5}^{3\sigma} < 0.29\times10^{-11}~{\rm photons\over cm^2~s}$\\
\hline
\end{tabular}
\caption[]{Observation log}
\end{table}

This paper reports on a 
multifrequency campaign to observe the EGRET-detected 
(v. Montigny et al. 1995) BL Lac object 
W~Comae (ON 231, 1219+285).  The source has a redshift of 0.102
and we therefore expect it to suffer a cutoff due to attenuation
somewhere in the TeV region.

\section{Observations}

W~Comae was observed quasi-simultaneously in February 1996 covering the 
entire range of the electromagnetic spectrum from GHz to TeV energies. 
The 
observatories, observation dates and fluxes are listed in Table 1.

The IUE campaign found W~Comae in an active state about two weeks before 
the bulk of the observations, organised around an EGRET observation. 
Unfortunately, due to solar angle constraints, UV observations
simultaneous with EGRET could not be scheduled. Optical 
monitoring at Tuorla showed that W~Comae was also unusually
bright in the optical at that time, whereas the GHz 
emission did not show enhanced activity.
A 
ToO observation with XTE was performed about a week later, but yielded 
only an upper limit indicating flux below previously reported levels. 
The optical campaign at McDonald Observatory showed no 
signs of extraordinary activity during the EGRET observations\footnote{
Monitoring at McDonald Obseratory was done for all flat-spectrum radio
sources in the FOV of EGRET.
None of these sources was in an unusually
bright optical or $\gamma$-ray state during the program. 
Radio monitoring showed that one source, 1222+216, was in a
fairly active state characterised by a rapidly increasing
high-frequency radio flux.  The source
was, however, not within the 
FOV of EGRET used in the
W~Comae observation.}. 
Optical
monitoring at Tuorla 
shows that the flux declined steadily from the time 
of the IUE observations to the time of the optical/$\gamma$-ray effort
(see also Tosti et al. 1997). 
The V-magnitude decreased by $\Delta m_{\rm V}
\sim 0.5$ during this time.
Our EGRET observations showed that W~Comae was weaker 
than during its strongest detection in November 1993 
by a factor of $1.5$, resulting in a detection above 100~MeV with just 
4.2$\sigma$. This weak signal does not allow for a determination of the 
spectral index. Since most counts were found at energies below 300~MeV, 
this indicates a steeper spectrum than observed before, i.e. a photon index of 
2.0 rather than 1.4 (v. Montigny et al. 1995) or maybe even a peak
in the spectrum in this energy range. 
Only upper limits were derived at TeV energies from 
HEGRA (Petry et al.~1997) and (preliminary) Whipple data analysis. 
Rapid blazar variability with outbursts on time scales of days or less has
been reported by
Wagner and Witzel (1995) and Gaidos et al. (1996).  It is difficult to
catch such short-duration outbursts
in multi-frequency campaigns.  
Our spectrum shown in Fig.1 averages the emission
over a few weeks. 
The constancy of the radio flux and the only small differences
in the optical fluxes between the time of the IUE and EGRET
observations
make us  quite confident
that the data
represent a realistic snapshot spectrum of
W~Comae.  

\section{SSC and PIC models }

Bearing in mind
the caveats mentioned above,
our quasi-simultaneous spectrum shown in Fig.1 constrains
emission models.  

Let us first consider a very simple SSC model, since it
is the most economic one requiring only the observed photons, electrons,
and a magnetic field.  
We adopt the simple picture 
where the $\alpha\sim 1$ IR-to-UV emission is optically thin synchrotron
emission from a homogenous, spherical blob in the jet.  
We adopt an IR spectral break at
$\epsilon_{\rm syn,b}\sim 0.1$~eV and a UV
cutoff at $\epsilon_{\rm syn,c}\sim 10$~eV, better spectral coverage
would be required to determine these energies more precisely.
Self-Compton scattering of the synchrotron spectrum produces a wider
and more rounded $\gamma$-ray spectrum between a smoother $\gamma$-ray
break energy
$\epsilon_{\rm ic,b}\sim \epsilon_{\rm syn,b}\gamma_{\rm b}^2$
and turnover energy
$\epsilon_{\rm ic,c}\sim \epsilon_{\rm syn,c}\gamma_{\rm c}^2$.
We put $\epsilon_{\rm ic,b}=100$~MeV, so that we obtain 
$\epsilon_{\rm ic,b}/\epsilon_{\rm syn,b}=
\gamma_{\rm b}^2\sim 10^9$ from the above equation.  
With the formula for the characteristic synchrotron energy
$\epsilon_{\rm syn}\sim \delta(B/B_{\rm c})\gamma^2 m_{\rm e}c^2$
where $\delta$ denotes the Doppler factor of the jet
and $B_{\rm c}=4\times 10^{13}$~G we obtain an expression for the
magnetic field, viz. $B\sim (\epsilon_{\rm syn}/\delta m_{\rm e}c^2)
B_{\rm c}\gamma^{-2}$.  Inserting $\epsilon_{\rm syn}=\epsilon_{\rm syn,b}$
and $\gamma^{-2}=10^{-9}$ this yields an estimate of the 
$B\sim 10^{-2}/\delta$~G.  The $\gamma$-ray turnover in this
simple model would be just
below $\epsilon_{\rm ic,c}\sim (\epsilon_{\rm syn,c}/
\epsilon_{\rm syn,b})^2\epsilon_{\rm ic,b}\sim 1$~TeV.  Thus,
within the uncertainties of this simple SSC model, one would expect
a rather low magnetic field strength and a $\gamma$-ray turnover just
below TeV.


\begin{figure}
\centerline{\psfig{figure=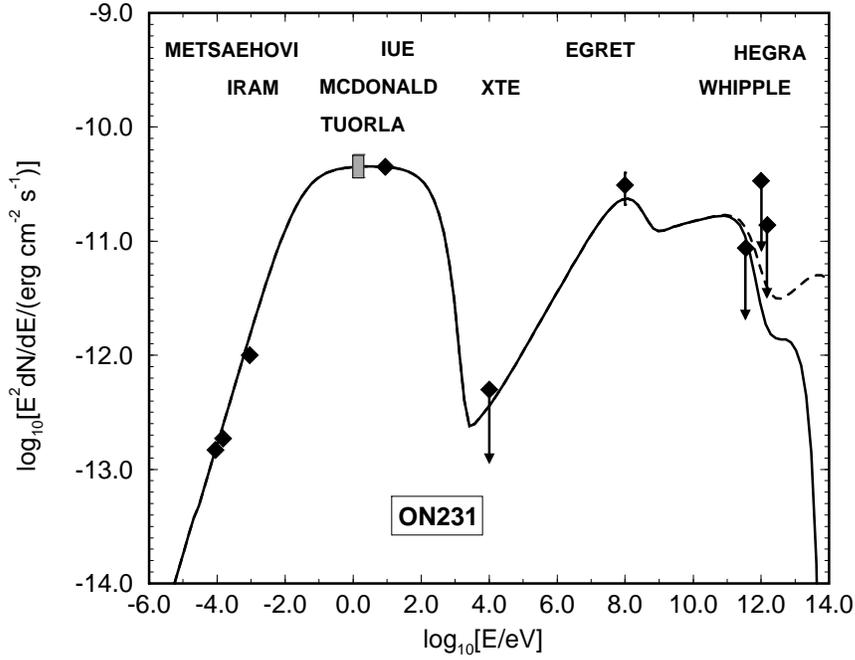,height=10cm,width=12cm}}
\caption[]{Quasi-simultaneuous multi-frequency spectrum of W Comae during
the period February 08 to March 10, 1996.  The flux range in the optical
indicates the observed variations 
between the time of the IUE and the EGRET observation.
For the XTE flux limit
and the differential EGRET flux
a photon index of 2, for the TeV  limits a photon index of 3 was 
assumed.
The {\it solid line} shows a
proton blazar model fit taking into account 
external $\gamma$-ray absorption (adopting $H_\circ=75$~km~s${-1}$~Mpc$^{-1}$,
$\Omega=1$, $\Lambda=0$, and the average EIB model from
MacMinn and Primack 1996).  The {\it dashed line} shows the model
spectrum without external absorption.}
\end{figure}

By analogy, the PIC model considers radiation induced by
protons scattering off the synchrotron photons in the jet.
Reaching much higher energies than the electrons, the protons 
produce
synchrotron $\gamma$-rays themselves (peak at
$\sim 100$~MeV in Fig.1) and initiate
electromagnetic cascades by photo-pair and photo-pion production
(bumpy spectrum $>100$~MeV).
Parameters for the model fit shown in Fig.1 are 
$\delta_{10}=3.6$, proton-to-electron cooling
rate ratio $\xi=10^{-3}$ (at resp. maximum energies),
proton-to-electron energy density ratio $\eta=7
\eta_{\rm CR}$ ($\eta_{\rm CR}=100$ denotes the value for local
cosmic rays),
jet opening-angle $\Phi=2^\circ$,
and  jet luminosity 
$L=8\times 10^{44}$~ergs~s$^{-1}$.  A magnetic field of $B=37$~G
is computed assuming equipartition.
External absorption does not have to be strong, but appears to
be necessary to avoid a conflict between the predicted flux and
the TeV limits.
By contrast, the SSC model seems to be consistent with
a cutoff just below TeV and therefore does not require any external
absorption.

\section{Discussion}

The $\gamma$-ray observations have detected W~Comae in a relatively low state 
of emission. Optical monitoring shows that the source was in decline during 
the $\gamma$-ray observations, supporting the picture of other 
multifrequency campaigns which found optical through $\gamma$-ray emission to 
be temporally associated. The question remains, however, 
why it has not been detected by Whipple or HEGRA. Extrapolating from the 
EGRET flux with an assumed photon index of 2.0, one exceeds the TeV limits, 
i.e. such a spectrum would have to 
steepen due to either internal or external absorption. 
SSC models with rather weak
magnetic fields of $B\sim 0.001-0.01$~G
can explain the HEGRA/Whipple 
non-detections with W~Comae being a member of the RBL class, which have 
intrinsically lower cutoffs than XBLs such as Mrk~421 (Maraschi et al. 1996). 
Detailed modeling will show whether the faintness of W~Comae in X-rays
is consistent with an SSC explanation of the $\gamma$-rays.

The PIC model (Mannheim 1993) predicts an approximate $E^{-2}$ spectrum
at EGRET energies and an approximate $E^{-3}$ spectrum
at TeV energies.  
For Mrk421, the $>$TeV spectrum has indeed been observed to be smooth 
up to 5-8~TeV and
to be steeper than the EGRET
spectrum (Petry et al.~1996, Krennrich et al.~1997).
This indicates that the cosmic pair-creation optical depth at
5~TeV and $z=0.03$ is very small.  Using a model for the EIB
based on the work of MacMinn and Primack (1996), we obtain
$\tau_{\gamma\gamma}(z=0.03,5~{\rm TeV})\sim 0.3$ and
$\tau_{\gamma\gamma}(z=0.10,1~{\rm TeV})\sim 0.8$.
The steepness of the
TeV spectrum of W~Comae predicted by PIC is not entirely
sufficient to reduce the predicted flux 
below the Whipple flux limit, although such a conclusion
may well be changed taking into account the 
flux variations seen in the optical over the observation campaign.
Taking into account the
above cosmic optical depth, we obtain marginal agreement 
of the model fit with the
data.  Another possibility is that absorption by
a warm dusty torus
(Protheroe and Biermann 1996)
is important.  

\acknowledgements
RDG was supported by DFG travel grant Az Sta 173/21-1. 
We thank D. de Martino for 
assistance during the IUE observations and acknowledge
the cooperative spirit of the EGRET team, in particular R. Hartman, as well
as the HEGRA and Whipple collaborations.

\end{document}